\documentclass[twocolumn,aps,amssymb,,superscriptaddress]{revtex4}
\usepackage{graphicx}

\begin{document}
\title{Scaling of THz-conductivity at metal-insulator transition in doped manganites}
\author{A. Pimenov}
\author{ M. Biberacher}
\author{ D. Ivannikov}
\author{ A. Loidl}
\affiliation{Experimentalphysik V, Center for Electronic
Correlations and Magnetism, Universit\"{a}t Augsburg,
86135 Augsburg, Germany} %
\author{A.$\>$A. Mukhin}
\affiliation{Experimentalphysik V, Center for Electronic
Correlations and Magnetism, Universit\"{a}t Augsburg,
86135 Augsburg, Germany} %
\affiliation{General Physics Institute of the Russian Acad. of
Sciences, 119991 Moscow, Russia}%
\author{Yu. G. Goncharov}
\affiliation{General Physics Institute of the Russian Acad. of
Sciences, 119991 Moscow, Russia}%
\author{A. M. Balbashov}
 \affiliation{Moscow Power Engineering Institute, 105835 Moscow, Russia}
\date{\today}

\begin{abstract}
Magnetic field and temperature dependence of the Terahertz
conductivity and permittivity of the colossal magnetoresistance
manganite Pr$_{0.65}$Ca$_{0.28}$Sr$_{0.07}$MnO$_3$ (PCSMO) is
investigated approaching the metal-to-insulator transition (MIT)
from the insulating side. In the charge-ordered state of PCSMO
both conductivity and dielectric permittivity increase as function
of magnetic field and temperature. Universal scaling relationships
$\Delta \varepsilon \propto \Delta \sigma$ are observed in a broad
range of temperatures and magnetic fields. Similar scaling is also
seen in La$_{1-x}$Sr$_x$MnO$_3$ for different doping levels. The
observed proportionality points towards the importance of pure
ac-conductivity  and phononic energy scale at MIT in manganites.
\end{abstract}

\pacs{71.30.+h,75.47.Gk,72.20.Ee,77.22.Ch} \maketitle


\textit{Introduction.} Physical properties of doped manganites are
governed by a complex interplay of charge, lattice, orbital and
spin degrees of freedom, which lead to a large variety of unusual
effects \cite{jaime}. While known since early works by Jonker and
van Santen \cite{jonker} and Wollan and Koehler \cite{wollan} the
interest in these compounds was enormously stimulated by the
observation of large magnetoresistance effects in thin manganite
films \cite{gmr}. It is generally accepted now \cite{dagotto01}
that the magnetoresistance effects in manganites are strongly
influenced by electronic phase separation \cite{nagaev} between
paramagnetic or antiferromagnetic (AFM) insulating and
ferromagnetic (FM) metallic regions which coexist on a microscopic
scale close to the metal-to-insulator transition (MIT).

Particularly large values of magnetoresistance have been observed
\cite{yoshizawa} in Ca-doped PrMnO$_3$ (PCMO) \cite{jirak}, which
has been explained as a consequence of the first-order transition
\cite{hardy,tomioka96} between charge- and orbitally-ordered
insulating \cite{zimmermann} and ferromagnetic metallic phases
\cite{baca}. Although phase-separation effects are still important
in PCMO \cite{dagotto01,anane,blake,fisher,uehara}, in order to
fully understand the MIT on the basis of electronic phase
separation a modification of the standard approach is necessary
\cite{burgy,hotta}. Additional fine-tuning of PCMO by the
substituition of Ca$^{2+}$ by Sr$^{2+}$  \cite{tomioka} leads to
eleven orders-of-magnitude changes in resistivity in magnetic
field \cite{raveau,jorg}. The subtle balance of various energy
scales in (Pr:Ca:Sr)MnO$_3$ results in new physical effects like
switching of the MIT by light, X-ray, or electric field
\cite{miyano,fiebig,kiryukhin,asamitsu}.

In this paper we present the results of the dynamic conductivity
experiments at THz frequencies on the insulating side of the
metal-to-insulator transition. On approaching the transition a
linear relationship between conductivity and dielectric
permittivity is observed, both using magnetic field and
temperature as tuning parameters. This linearity points towards
the importance of the characteristic frequency scale  of few THz
for the charge dynamics. Closely similar scaling behavior can be
stated in Sr-doped LaMnO$_3$, too.

\textit{Experimental details.} Single crystals of
Pr$_{0.65}$Ca$_{0.28}$Sr$_{0.07}$MnO$_3$ (PCSMO) which were used
in the present experiments were grown by the floating-zone method
with radiation heating \cite{preparation}. The samples were
characterized using various experimental techniques \cite{jorg}
and the B-T phase diagram has been constructed. Crystals with a
typical growth direction [100] were cut from the main rod to a
thickness of 0.1-1\,mm.

On cooling from room temperature, in zero magnetic field PCSMO
reveals a structural phase transition at $T_{CO}\simeq 210$\,K
into the charge-ordered (CO) insulating phase. On further cooling
an AFM phase transition is observed at $T_{\rm N}\simeq 170$\,K
and below $T_{irr} \simeq 100$\,K strong hysteresis effects in
magnetic fields are observed. Without magnetic field the samples
remain in the insulating state down to the lowest temperatures.
Strong magnetoresistance effects as high as $10^{10}$ at $T=50$\,K
are observed  below $T_{CO}$. Typical values of the external
magnetic field to switch between CO insulating and FM metallic
state are in the range 2-5 T. Detailed discussion of these results
is given in Ref. \cite{jorg}. The B-T phase diagram for the
samples under investigation \cite{jorg} agrees well with published
results \cite{tomioka,tomioka96}.

The dynamic conductivity experiments for frequencies $0.1$\,THz$\,
<\nu <1 $\,THz were carried out in a Mach-Zehnder interferometer
\cite{volkov,magnetite} which allows the measurements of
transmittance and phase shift of a plane-parallel sample. PCSMO
samples of different thicknesses have been utilized. To ensure
mechanical stability the thinnest sample ($\sim 0.1$\,mm thick)
was glued onto a MgO substrate. The electrodynamic properties of
the substrate were obtained in a separate experiment. The
experimental data for both, single layer and two-layers systems
have been analyzed using the Fresnel optical formulas for the
complex transmission coefficient \cite{born}. The absolute values
of the complex conductivity $\sigma ^{*}=\sigma _{1}+i\sigma _{2}$
and dielectric permittivity
$\varepsilon^*=\varepsilon_1+i\varepsilon_2=\sigma^*/i\varepsilon_0\omega$
were determined directly from the measured spectra. Here
$\varepsilon_0$ and $\omega=2\pi\nu$ are the permittivity of
vacuum and the angular frequency, respectively. The experiments in
external magnetic fields were performed in a superconducting
split-coil magnet, which allowed to carry out transmission
experiments in magnetic fields up to 7\,T. In order to exclude the
influence of the Faraday effect \cite{zvezdin}, the experiments
were carried out in external magnetic fields perpendicular to the
propagation of the electromagnetic beam (Voigt geometry).

\begin{figure}[]
\includegraphics[width=8cm,clip]{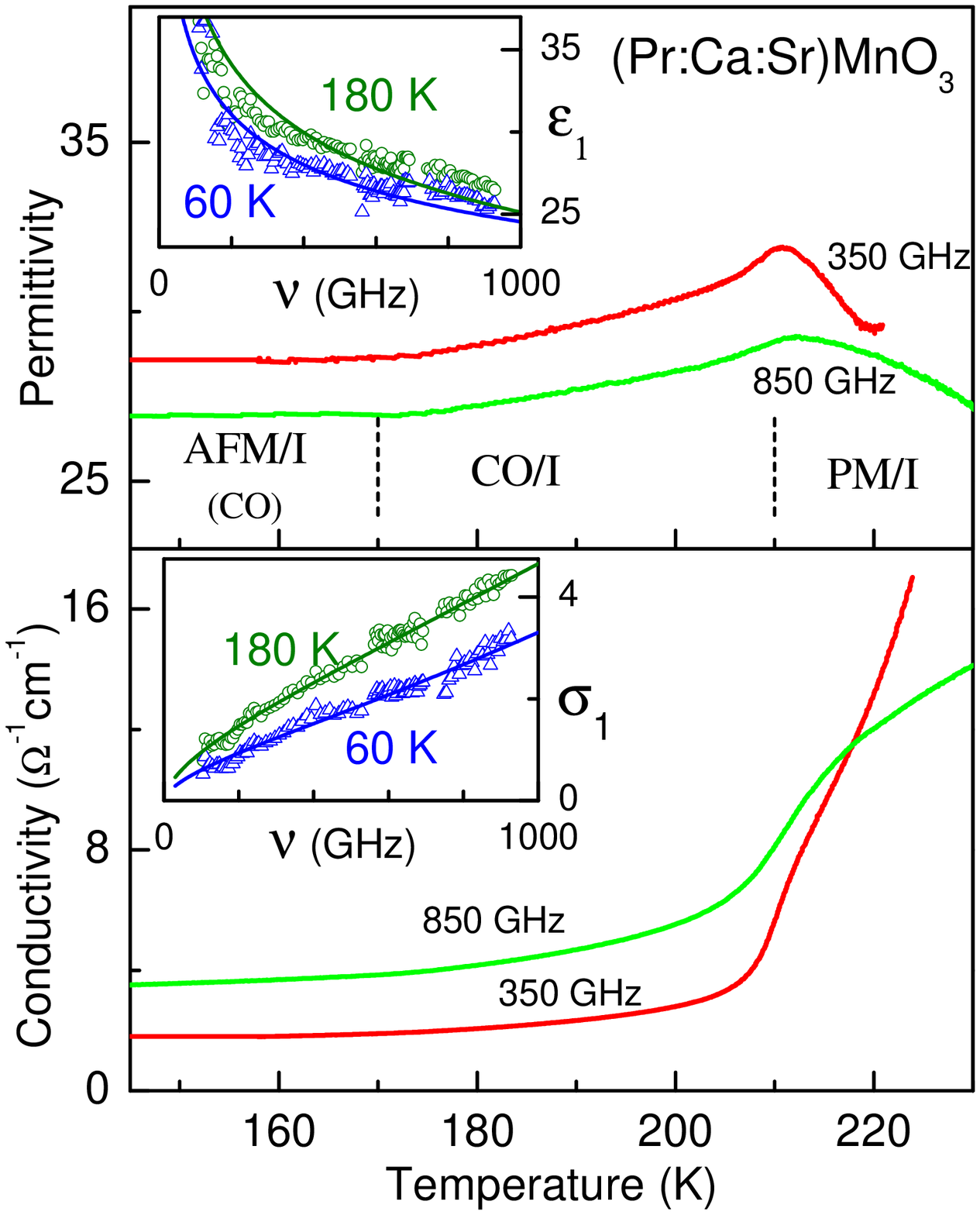}
\caption{(color online) Temperature dependence of the permittivity
(upper panel) and conductivity (lower panel) of
Pr$_{0.65}$Ca$_{0.28}$Sr$_{0.07}$MnO$_3$ at 350 GHz and 850 GHz.
Dashed lines indicate the transition temperature between different
phases: PM - paramagnetic, CO - charge ordered, AFM -
antiferromagnetic, I - insulator. The insets show the frequency
dependencies of conductivity and permittivity in the THz frequency
range. Symbols correspond to the experimental data, lines are fits
according to Eq.\,(\ref{equniv}) and Ref. \cite{eps} using
$s=0.8$, $\nu_c=2.4\ $THz, $A\nu_c^s = 6.3\ \Omega^{-1}$cm$^{-1}$
(60 K), and $A\nu_c^s = 8.4\ \Omega^{-1}$cm$^{-1}$ (180 K).}
\label{ftdp}
\end{figure}

\textit{Results.} Figure \ref{ftdp} shows the temperature
dependence of the dynamic conductivity (lower panel) and
permittivity (upper panel) of PCSMO at Terahertz frequencies in
zero external magnetic field. The high-frequency properties reveal
only a weak temperature dependence below $T_{CO}\simeq 210$\,K.
The small changes of the conductivity in this temperature range
are in marked contrast to the activated behavior of the
dc-resistivity \cite{jorg}. This is the consequence of the hopping
mechanism of the charge transport and is typical for systems with
localization \cite{boettger}. For these systems it is generally
observed that the dc conductivity reveals an exponential
temperature dependence, while the temperature dependence of ac
conductivity is comparatively weak \cite{long,peter}. Approaching
$T_{CO}\simeq 210\ $K from below, the dynamic conductivity starts
to increase rapidly and the permittivity exhibits a cusp-like
maximum at $T_{CO}$. Similar to
 recent results on the MIT in Fe$_3$O$_4$
\cite{magnetite} the decrease of the permittivity for $T>T_{CO}$
can be ascribed to the growth of a Drude-like contribution of the
mobile carriers to the dynamic conductivity, which now is
dominated by dc contribution, while ac processes become reduced.

The insets in Fig.\,\ref{ftdp} show the  characteristic
conductivity and permittivity spectra of ${\rm PCSMO}$ in the
frequency range of our experiment. The conductivity (lower panel)
is an increasing function of frequency and follows well the
power-law dependence
\begin{equation}\label{equniv}
    \sigma_1(\nu) = \sigma_{dc}+\frac{A
    \nu^s}{1+(\nu/\nu_c)^4}
\end{equation}
Here $\sigma_{dc}$ is the dc-conductivity and $s$ is the power-law
exponent. Compared to the conventional expression \cite{jonsher}
the additional frequency cutoff $\nu_c$ has been introduced in
order to preserve the finiteness of the conductivity spectral
weight. The application of the Kramers-Kronig transformation to
$\sigma(\nu)$ leads to an expression for the frequency dependence
of the dielectric permittivity \cite{magnetite,eps}, which has
been used to fit the spectra in the upper inset of
Fig.\,\ref{ftdp}. Comparatively poorer agreement between
experimental data and the calculations are due to small relative
changes in the dielectric permittivity ($\sim 25\% $ in our
frequency range). By investigating PCSMO using broadband
dielectric spectroscopy a superlinear power law has been detected
\cite{peter} at high frequencies and low temperatures. The
downward curvature in $\varepsilon_1(\nu)$ above 800 GHz (upper
inset in Fig.\,\ref{ftdp}) probably corresponds to this regime.

\begin{figure}[]
\includegraphics[width=7.5cm,clip]{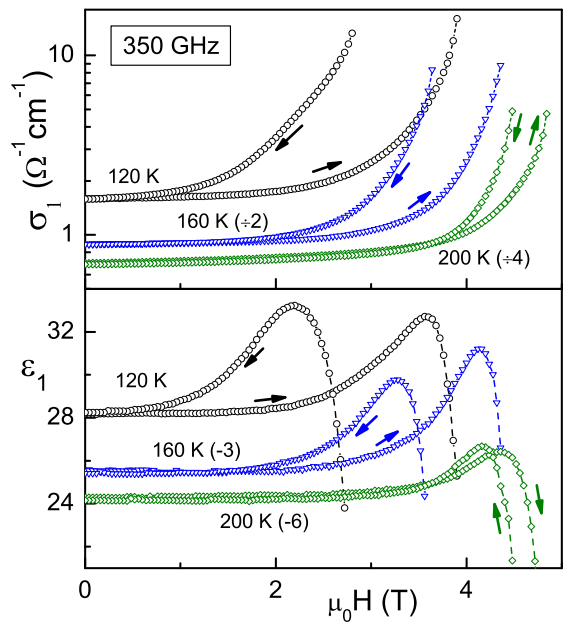}
\caption{(color online) Magnetic field-dependence of the
permittivity (lower panel) and conductivity (upper panel) of
Pr$_{0.65}$Ca$_{0.28}$Sr$_{0.07}$MnO$_3$ at 350 GHz and for
different temperatures. The curves were shifted for clarity as
indicated in the brackets. } \label{f350}
\end{figure}

Figure \ref{f350} shows the magnetic field-dependence of the
conductivity and the dielectric permittivity of PCSMO at
$\nu=350\,$GHz as a function of the external magnetic field.
Starting from $B=0$ and for increasing fields both, the
conductivity and permittivity initially increase. The increase of
the dielectric permittivity approaching the MIT in doped
semiconductors has been investigated earlier and is known as
\textit{dielectric catastrophe} \cite{mott}. For even higher
magnetic fields the increase of the conductivity in PCSMO
continues, but the dielectric permittivity starts to decrease.
Interestingly, similar behavior is observed in Fig.\,\ref{ftdp}
with temperature as a tuning parameter approaching the MIT. Based
on structural data, the temperature range close to the MIT has
been identified as a melting of the charge-ordered state in PCSMO
\cite{tomioka,yoshizawa}. In analogy, the upturn in the magnetic
field-dependence of the dielectric permittivity in
Fig.\,\ref{f350} is probably due to the melting of the
charge-ordered state under the influence of the external magnetic
field.

\begin{figure}[]
\includegraphics[width=8.5cm,clip]{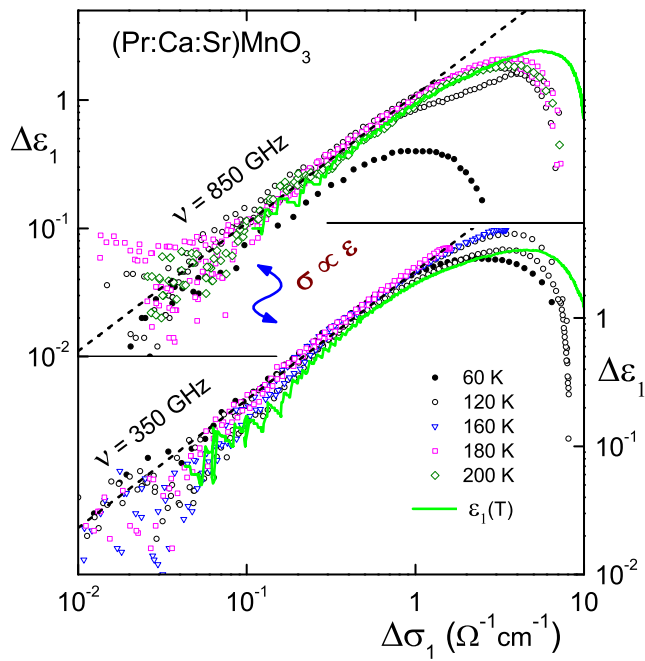}
\caption{(color online) Dielectric permittivity of PCSMO
re-plotted as function of conductivity at $\nu = 350\,$GHz (lower
panel) and $\nu = 850\,$GHz (upper panel). The data are shown for
fixed temperatures varying the magnetic field (symbols) and for
$B=0$ varying the temperature (solid lines). The dashed lines
indicate linear scaling between permittivity and conductivity.}
\label{fscal}
\end{figure}

Figure \ref{fscal} shows the dielectric permittivity of PCSMO
re-plotted as function of the conductivity. This plot with
magnetic field as a parameter allows to check a possible scaling
between conductivity and dielectric permittivity. In this
presentation, the zero-field values of $\sigma_1$ and
$\varepsilon_1$ have been subtracted and only the field-dependent
changes are shown. Surprisingly, all data between 60\,K and
200\,K, i.e in the charge-ordered state, converge into one
universal linear relationship of permittivity and conductivity
which is emphasized by the dashed lines in Fig.\,\ref{fscal}.
Solely close to the melting of the CO state this proportionality
between conductivity and permittivity breaks down. We note that
the same proportionality between $\sigma_1$ and $\varepsilon_1$ is
observed with the temperature as explicit parameter. This data
corresponds to the temperature dependence in Fig.\,\ref{ftdp} and
are shown in the scaling plot (Fig.\,\ref{fscal}) as solid line.

\textit{Discussion.} An explanation for the scaling $\sigma_1
\propto \varepsilon_1$ can be obtained on the basis of the
power-law of the conductivity, Eq. (\ref{equniv}). Within this
model the dielectric permittivity can be proportional to the
conductivity neglecting $\sigma_{dc}$  as well as
$\varepsilon_{\infty}$ and assuming fixed power-law exponent $s$
and cutoff frequency $\nu_c$ \cite{eps}. This constitutes an
important result of this work: dielectric constant and
conductivity in manganites are dominated solely by ac
contributions when approaching the MIT. Taking the power-law
exponent $s=0.8$ from the fits in Fig.\,\ref{ftdp} and using the
scaling of Fig.\,\ref{fscal} the cutoff frequency can be estimated
as $\nu_c \approx 2\,$THz, which is in agreement with $\nu_c =
2.4\,$THz from the fits in Fig.\,\ref{ftdp}. (We note that
changing the value of the power-law exponent $s$ only weakly
influences the cutoff frequency).  The cutoff frequency of the
insulating state lies in the frequency range of phonons and points
towards the phononic mechanism of the hopping conductivity in
PCSMO.

At the same time, dc and ac conductivity are closely connected
\cite{boettger} especially close to MIT. Depending upon the
conduction mechanism, this can be due to hopping mechanism of the
conductivity \cite{dyre}, or reflect different frequency scales in
the phase-separation scenario \cite{dagotto01}. In the former case
the observed cutoff frequency characterizes the maximum hopping
rate of charge carriers, which is therefore phonon-mediated.
Within the phase-separation scenario \cite{dagotto01} the cutoff
characterizes the relaxation within an elementary RC-circuit
between insulating and metallic droplets.


\begin{figure}[]
\includegraphics[width=8.5cm,clip]{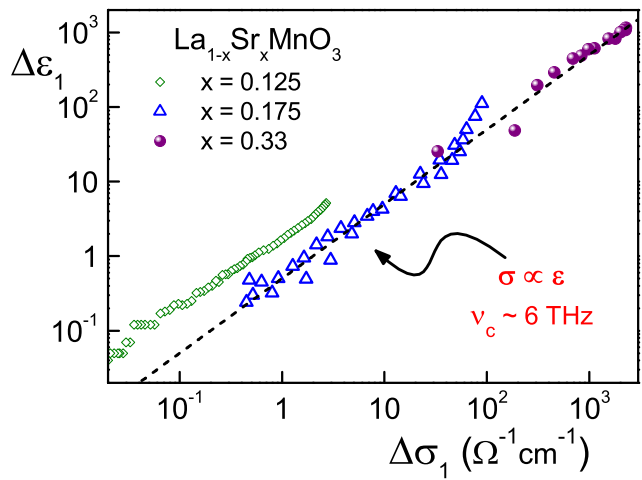}
\caption{(color online) Dielectric permittivity of
La$_{1-x}$Sr$_x$MnO$_3$ plotted as function of conductivity with
the temperature as tuning parameter and for $B=0$. Open diamonds -
$x=0.125$ (single crystal, \cite{localiz}), open triangles -
$x=0.175$ (single crystal, \cite{localiz}), closed circles -
$x=0.33$ (thin film, \cite{hartinger}). The data correspond to the
frequency range $400-1000\,$GHz. The dashed line indicates a
linear scaling between permittivity and conductivity.}
\label{flsmo}
\end{figure}

In contrast to recent reports on colossal dielectric constants in
manganites \cite{colossal} the observed high-frequency values are
comparatively low. Neither contacts nor internal-barrier layers
contributions \cite{contacts} play a role in THz quasi-optic
experiments. However, due to the dominating term $\varepsilon_1
\sim (\sigma_1-\sigma_{dc})/\nu \sim \nu^{s-1}$\cite{eps}, the
permittivity at low frequencies indeed can diverge, and an
estimate with the parameter in Fig.\ \ref{ftdp} leads to the
values above $\varepsilon_1 \sim 1000$ at kHz frequencies.

Finally, we compare the observed scaling behavior in PCSMO with
the temperature dependence of the conductivity and permittivity in
Sr-doped LaMnO$_3$. The parent compound LaMnO$_3$ is an insulator
and orders antiferromagnetically below $T_N = 140\,$K. On
substituting La$^{3+}$ by Sr$^{2+}$
\cite{urushibara,localiz,seeger} metallic behavior sets in for
doping levels above $\sim 16 \%$. Figure \ref{flsmo} shows the
parametric dependence of the dielectric permittivity in
La$_{1-x}$Sr$_x$MnO$_3$ for different doping levels. The data
correspond to two metallic compositions, $x=0.175$ \cite{localiz}
(single crystals) and $x=0.33$ \cite{hartinger} (thin films), and
to one composition in the insulating part of the diagram,
$x=0.125$ \cite{localiz} (single crystals). In full analogy with
the results for PCSMO, the data in Fig. 4 reveal a direct
proportionality between conductivity and permittivity. Again,
similar arguments allow to estimate the characteristic frequency
of the underlying process, $\nu_c \sim 6\,$THz. This documents
that at high frequencies ($\nu > 300\ $GHz) even in the "metallic
manganites" the ac conductivity dominates over dc processes.

\textit{Conclusions.} Dynamic conductivity and permittivity of
(Pr:Ca:Sr)MnO$_3$ have been investigated close to the magnetic
field-induced metal-to-insulator transition (MIT) and at THz
frequencies. Approaching the transition from the insulating side
the changes in conductivity and permittivity are directly
proportional to each other. Within simple arguments this
proportionality indicates that the charge transport close to MIT
is governed by ac processes with a characteristic frequency in the
phonon range. Closely similar scaling can be observed in Sr-doped
LaMnO$_3$ thus suggesting the universality of the MIT in
manganites.

The stimulating discussion with P. Lunkenheimer is gratefully
acknowledged. This work was supported by BMBF(13N6917/0 - EKM) and
by DFG (SFB 484).


\begin{thebibliography}{99}
\bibitem{eps} Application of the Kramers-Kronig
analysis to Eq.\,(\ref{equniv}) gives the following expression for
the dielectric constant:
\begin{eqnarray*}
&& \varepsilon_1(\nu)-\varepsilon_{\infty}= -\frac{\sigma_2(\nu)}{
2 \pi \nu \varepsilon_0} =\frac{\sigma_1(\nu) - \sigma_{dc}}{ 2
\pi \nu \varepsilon_0} \Bigg\{ \tan(s\pi/2) -  \\
&& -  \frac{(\nu/\nu_c)^{1-s}}{\cos(s\pi/2)}
 \left[\sin[(s+1)\frac{\pi}{4}]+\left(\frac{\nu}{\nu_c}\right)^2\cos[(s+1)\frac{\pi}{4}]\right]
\Bigg\} \quad , \quad
\end{eqnarray*}
where $\varepsilon_{\infty}$ is the high-frequency dielectric
constant. This expression is valid both for $s<1$ and for $s>1$.
\end{thebibliography}
\end{document}